\documentclass[prl,superscriptaddress,twocolumn,showpacs,preprintnumbers,amsmath,amssymb]{revtex4}

\usepackage{graphicx}% Include figure files
\usepackage{dcolumn}% Align table columns on decimal point
\usepackage{bm}% bold math

\begin{document}

\title{Helicoidal magnetic order in a clean copper oxide spin chain compound}

\author{L.~Capogna, M. Mayr, P. Horsch, M. Raichle,
R.K.~Kremer, M.~Sofin, A.~Maljuk, M.~Jansen, and B.~Keimer}

\address{Max-Planck-Institut f$\ddot u$r Festk$\ddot o$rperforschung,
Heisenbergstrasse 1 Stuttgart D-70569, Germany \\ } \date{\today}

\begin{abstract} We report susceptibility, specific heat, and neutron
diffraction measurements on NaCu$_2$O$_2$, a spin-1/2 chain compound
isostructural to LiCu$_2$O$_2$, which has been extensively
investigated. Below 12 K, we find a long-range ordered, incommensurate
magnetic helix state with a propagation vector similar to that of
LiCu$_2$O$_2$. In contrast to the Li analogue, substitutional disorder
is negligible in NaCu$_2$O$_2$. We can thus rule out that the helix is
induced by impurities, as was claimed on the basis of prior work on
LiCu$_2$O$_2$.  A spin Hamiltonian with frustrated longer-range exchange
interactions provides a good description of both the ordered state and
the paramagnetic susceptibility.
\end{abstract}

\maketitle

Copper oxides are excellent model systems for low-dimensional
spin-1/2 quantum antiferromagnets. In particular, copper oxides
with magnetic backbones comprised of chains of CuO$_4$ squares
have been shown to exhibit quasi-one-dimensional behavior. Two
classes of copper oxide spin chain materials are known. Compounds
in which adjacent squares share their corners are excellent
realizations of the one-dimensional (1D) spin-1/2 Heisenberg
Hamiltonian \cite{Kojima, Motoyama, Ami}. Linear Cu-O-Cu bonds along the spin chains give rise
to a large antiferromagnetic nearest-neighbor exchange coupling.
In compounds built up of edge-sharing squares, on the other
hand, the Cu-O-Cu bond angle is nearly 90$^\circ$, so that the
nearest-neighbor coupling is more than an order
of magnitude smaller ~\cite{Mizuno}. Because of the anomalously small
nearest-neighbor coupling, longer-range frustrating exchange
interactions have a pronounced influence on the physical
properties of these materials. Edge-sharing copper oxides thus
provide uniquely simple model systems to test current theories of
spin correlations in frustrated quantum magnets.

At low temperatures, the ground state of edge-sharing copper oxides
is either a 3D-ordered antiferromagnet \cite{Sapina, Fong, Chung} or a spin-Peierls state \cite{Hase},
depending on whether interchain exchange interactions or spin-phonon
interactions are dominant. In the former case, the magnetic order is
almost always collinear. An interesting exception was recently
discovered in LiCu$_2$O$_2$ \cite{Zvyagin, Choi, Masuda}, which undergoes a transition to a
magnetic helix state at low temperatures. While such a state is
expected for classical spin models with frustrating interactions,
quantum models predict a gapped spin liquid state in the range of
exchange parameters that was claimed to describe the spin system in
LiCu$_2$O$_2$. Since the ionic radii of Li$^+$ and Cu$^{2+}$ are
similar, chemical disorder was identified as a possible solution to
this puzzle.  Indeed, a chemical analysis of the sample used in the
neutron scattering study of Ref.~\cite{Masuda} showed that about 16\%
of the Cu$^{2+}$ ions in the spin chains were replaced by nonmagnetic
Li$^+$ impurities. Since even much lower concentrations of nonmagnetic
impurities are found to induce magnetic long-range order in other
quasi-1D spin-gap systems, the authors of Ref.~\cite{Masuda}
attributed the unexpected helix state to the highly disordered lattice
structure of LiCu$_2$O$_2$.

Here we report magnetic susceptibility, specific heat, and neutron
diffraction data on NaCu$_2$O$_2$, a Mott insulator that is
isostructural to LiCu$_2$O$_2$. However, since Na$^+$ is much
larger than Cu$^{2+}$, substitutional disorder is {\it a priori}
unlikely in NaCu$_2$O$_2$. Chemical analysis and neutron
diffraction data confirm that Na-Cu inter-substitution is
negligible in our NaCu$_2$O$_2$ samples. Below a N\'{e}el
temperature $T_{\rm N}$ of 12K, we find an incommensurate helix
state similar to that in LiCu$_2$O$_2$.
Contrary to the interpretation advocated in Ref.~\cite{Masuda},
this state is thus the ground state of a spin chain system without
impurities. The incommensurability along the chain axis is such
that the magnetic unit cell is nearly quadrupled with respect to
the chemical cell.  Together with an analysis of the
susceptibility data, this indicates that an antiferromagnetic
next-nearest-neighbor interaction along the spin chain is the
dominant exchange interaction in NaCu$_2$O$_2$. This disagrees
with the spin Hamiltonian recently proposed for LiCu$_2$O$_2$
\cite{Masuda}. Each copper oxide chain thus contains two
interpenetrating, but nearly independent 1D spin systems. 
A model with longer-range exchange interactions provides a
quantitative description of the ground state and paramagnetic susceptibility of
this spin-1/2 system.

Micro-crystalline powder samples of NaCu$_2$O$_2$ were synthesized
via the azide/nitrate route in specially designed containers
~\cite{Trinschek, Sofin, Tams}. The starting materials were milled,
pressed in pellets under $10^5 \rm N$, dried in vacuum (10$^{-3}$
mbar) at $150^\circ$ C for 12 h, and placed in an argon atmosphere
in a tightly closed steel container provided with a silver inlay.
Finally, in a flow of dry argon the following temperature sequence
was applied: 25 to $260 ^\circ$ C (100 $^\circ$C/h); 260 to $380 ^\circ$ C
(5 $^\circ$C/h), 380 to $500 ^\circ$C (50 $^\circ$C/h) with subsequent annealing
for 30 hours at $500 ^\circ$ C.  Powder X-ray diffraction patterns
for initial characterization were collected with a Stoe STADI-P
diffractometer using CuK$_{\alpha 1}$ and MoK$_{\alpha 1}$ radiation. No impurity phases
were detected, except small traces of CuO and Cu$_2$O. The single
crystals were grown by the self-flux method in an argon atmosphere
in platinum crucibles ~\cite{Maljuk}. The crystals have a platelet
shape with typical sizes of $7\times 3$ mm$^2$ and thickness of up
to 100 ${\mu}m$. X-ray diffraction from single crystals ground to
powder showed no sign of impurity phases within the resolution
limit. In addition, no platinum impurities were detected in our
crystals by inductively-coupled plasma atomic emission
spectroscopy (ICP-AES) measurements. Finally, we have used ICP-AES
to establish the chemical composition of our crystals as Na$_{1.00
\pm 0.02}$Cu$_{2.00 \pm 0.02}$O$_y$. The Na:Cu ratio is thus
identical to the ideal stoichiometry within the experimental
error. This should be compared to the samples studied in
Ref.~\cite{Masuda} whose chemical composition was quoted as
Li$_{1.16}$Cu$_{1.84}$O$_{2.01}$.

\begin{figure}
\centerline{\includegraphics[width=8.0cm,angle=0]{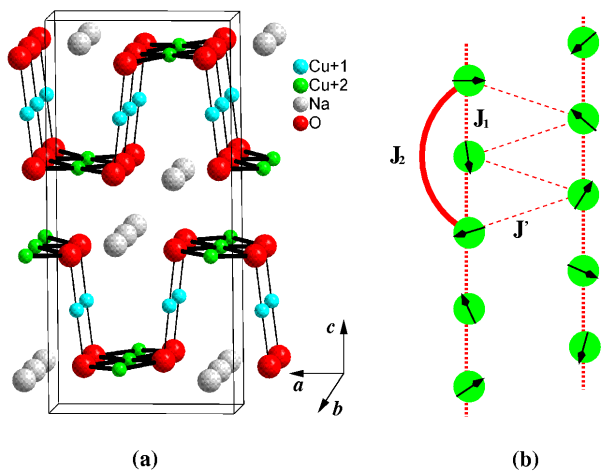}}
\caption{(a) Crystal structure, and (b) proposed magnetic
structure of NaCu$_2$O$_2$. In panel (b), the most important
magnetic interactions within a chain pair are sketched.} 
\label{LCfig1.eps}
\end{figure}

For the neutron diffraction measurements, a powder sample of
weight $\sim5$ g was fitted into a cylindrical vanadium container
and loaded in a standard helium-flow cryostat. High-resolution
diffraction patterns were collected at 3.5, 23 and 300 K on the
two-circle powder diffractometer D2B at the Institut
Laue-Langevin, Grenoble. A wavelength $\lambda=1.5946 \rm {\AA}$
was selected by a Ge[355] monochromator. The diaphragms were set
to 70/70 and the collimations to 10'-10'-10' before the
monochromator, between monochromator and sample, and between
sample and detector, respectively. Complementary neutron
diffraction data were taken on the high-flux powder diffractometer
D20 using a wavelength $\lambda=2.4 \rm {\AA}$. The diaphragms
were set to 70/10. Measurements of the magnetic susceptibility and
the heat capacity were carried out on powder samples as well as on
single crystals. The powder measurements were taken on a sample of
total mass 115 mg in a SQUID magnetometer (MPMS, Quantum Design).

A Rietveld analysis of the diffraction pattern at room temperature
confirms that NaCu$_2$O$_2$ has an orthorhombic crystal structure
(space group Pnma) with lattice parameters $a =6.2087$\rm {\AA},
$b=2.9343$\rm {\AA} and $c=13.0548$\rm {\AA}, in agreement with an
X-ray investigation of small single crystals.  A pictorial
representation of the lattice structure is shown in Fig. 1. The unit
cell of NaCu$_2$O$_2$ contains four magnetic Cu$^{2+}$ ions belonging
to two pairs of copper oxide chains running along $b$. The Cu-O-Cu
bond angle is $92.9^\circ$, somewhat larger than $87.2^\circ$, the
equivalent angle in LiCu$_2$O$_2$. The chains are separated from each
other in the $a$-direction by rows of non-magnetic Cu$^{+}$ ions, and
in the $c$-direction by Na$^{+}$ ions. Since the two chains within one
pair are shifted relative to each other by half a unit cell along $b$,
each chain can also be viewd as a single zig-zag chain
\cite{White}. The nearest-neighbor Cu$^{2+}$-Cu$^{2+}$ distances
between different chain pairs are considerably larger than those
within a pair. It is thus reasonable to expect that the system is
magnetically one-dimensional.

In order to check for possible Na-Cu inter-substitution, we have
refined the occupancies of Na and Cu sites in the Rietveld
analysis of the neutron diffraction data and obtained full
occupancy of both positions within an error of 2\%. This finding
is expected based on the different ionic radii of Na$^+$ and
Cu$^{2+}$, and it confirms the results of the chemical analysis
above. In contrast to LiCu$_2$O$_2$, substitutional disorder on
the copper oxide chains is therefore negligible in NaCu$_2$O$_2$.

Fig. 2 shows the uniform magnetic susceptibility of a NaCu$_2$O$_2$
powder as a function of temperature. The main feature of the curve is
a broad maximum at 52 K, which is characteristic of low-dimensional
spin systems and indicates a crossover to a state with
antiferromagnetic short-range order. A Curie-Weiss fit of the
high-temperature susceptibility for 200 $<T<$ 300 K yields a negative
Curie-Weiss temperature of $\Theta_{CW}=-62$ K, indicating predominant
antiferromagnetic interactions.  Because of the intrinsic 1D nature of
our system, we have fitted the susceptibility curve to the exact
solution of the $S=1/2$ Heisenberg chain with a single
antiferromagnetic interaction $J_{2}$ \cite{Johnston}, resulting from
the dominant Cu-O-O-Cu exchange path.  We added a diamagnetic
contribution from the closed atomic shells, which we estimate as
-52$\times$10$^{-6}$cm$^3$/mol from Pascal's increments
\cite{Selwood}.  The fit yields $J_{2}= 85$ K ($k_B$$\equiv$1) and
$g=2.07$, and the result is shown in Fig. 2. Our susceptibility data
are thus in good quantitative agreement (except small deviations) with
a model including a single antiferromagnetic interaction parameter
along the spin chains (Fig.  2), supporting the view of a single chain
as two interpenetrating, weakly coupled $J_2$-Heisenberg chains.

The low-temperature susceptibility and specific heat data plotted in
Fig. 3 indicate two magnetic phase transitions at 12 and 8 K.  Both
transition temperatures are much smaller than $J_2$, as expected based
on the quasi-one-dimensionality of the magnetic lattice. The
temperature dependent susceptibility of a single-crystal sample in the
three crystallographic directions is plotted in Fig. 3a. The
susceptibility in the $b$-direction, $\chi_b$, is not strongly
affected by the 12 K transition, while $\chi_a$ and $\chi_c$ are
suppressed. Below the 8 K transition, on the other hand, $\chi_a$ and
$\chi_b$ exhibit an upturn, reflecting the role of anisotropic
interactions. The specific heat anomaly at the 8 K transition is
weaker and obliterated by a modest magnetic field of 9T (Fig. 3b),
whereas the 12 K transition is more robust. Taken together, these data
suggest that the 8 K transition arises from spin canting. A similar
canting transition has been observed in Li$_2$CuO$_2$ \cite{Chung}.

\begin{figure}
\centerline{\includegraphics[width=8.0cm,angle=0]{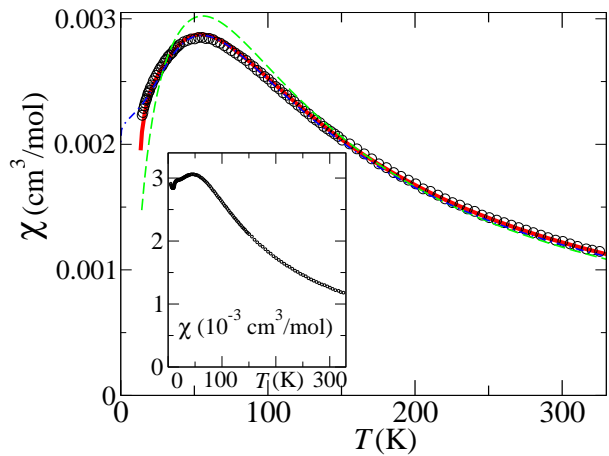}}
\caption{Temperature dependence of the static magnetic
susceptibility of a powder sample of NaCu$_2$O$_2$ in a magnetic
field B=0.1 T (open circles). The thick red line is the result of
a fit to the helix model described in the text, whereas the blue
line (dot-dashed, mostly hidden by the red one) is the (exact)
susceptibility of the 1D-Heisenberg chain ($J_2$=85K)
\cite{Johnston}. For comparison, the green line represents the
result of a fit to a simpler $J_1$-$J_2$ helix model (see text).
The inset shows the actually measured data, from which a Curie
contribution $\chi_{Cur}$=0.0105/T has been subtracted to obtain
the data presented in the main panel. Note that the Curie tail is
very weak, providing testimony to the high quality of the sample.}
\label{LCfig2.eps}
\end{figure}

These data indicate that the magnetic ground state is more complex
than suggested by the initial analysis of the paramagnetic
susceptibility. In order to determine the spin configuration in the
ground state, neutron powder diffraction data were taken on the
high-flux powder diffractometer D20. The results are shown in
Fig. 4. Several additional Bragg reflections are observed below 12 K
at low scattering angles. Since no high-angle counterparts of these
reflections are observed, and the intensities vanish at the magnetic
ordering temperature (inset in Fig. 4), these reflections can be
identified as magnetic. They can be indexed based on the magnetic
propagation vector $(0.5,\zeta,0)$ with $\zeta=0.227$, which
corresponds to a pitch angle $\phi_b=81.7^\circ$ along the
$b$-axis. Since an amplitude modulation of the copper spin is not
expected in a Mott insulating state, the intensities were compared to
models in which the spins form a helicoidal state with identical
amplitude on every Cu$^{2+}$ site, analogous to that observed in
LiCu$_2$O$_2$. An excellent refinement of the peak intensities (solid
line in Fig. 4) was obtained by assuming a helix polarized in the
$a$$b$-plane (Fig. 1), similar to LiCu$_2$O$_2$ \cite{Masuda,
comment}. The ordered moment per Cu$^{2+}$ ion at base temperature is
$ 0.56(4) \mu_B$.

\begin{figure}
%\centerline{\epsfysize=8cm \epsfbox{LCfig3.eps}}
\centerline{\includegraphics[width=8.0cm,angle=0]{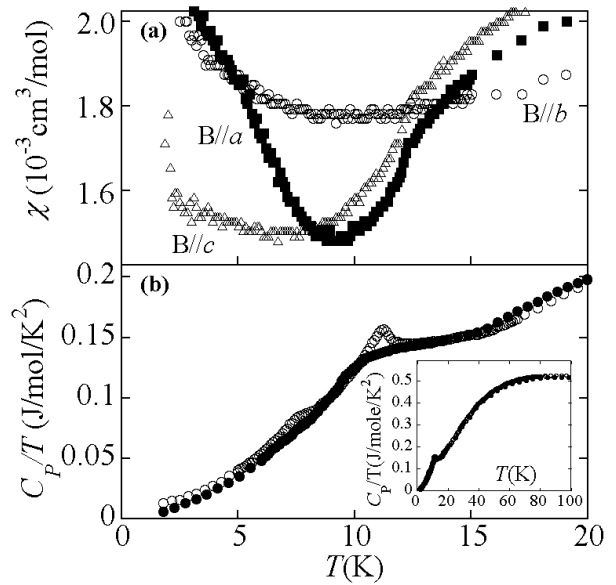}}
% preprint above, two-column below
%\centerline{\epsfysize=6cm \epsfbox{layout.eps}}
\caption{Single-crystal data for NaCu$_2$O$_2$ as a function of
temperature. (a) Susceptibility taken with a magnetic field of
0.1T applied along the principal crystallographic axes. (b)
Specific heat for two applied magnetic fields: 0T (open circles)
and 9T (full circles). The inset extends the temperature range.}
\label{LCfig3.eps}
\end{figure}

Since the observed helical spin structure of NaCu$_2$O$_2$ cannot be
accounted for within a simple nearest-neighbor Heisenberg model, we
reanalyzed the paramagnetic susceptibility in terms of a spin
Hamiltonian with longer-ranged interactions $J_d$. Such additional
couplings are expected for Cu-O-O-Cu bridges in edge-sharing copper
chains \cite{Mizuno}. We consider magnetic couplings up to a distance
$d$=4$b$. Whereas couplings $J_{d\geq2}$ are always antiferromagnetic,
$J_1$ can be either ferromagnetic or antiferromagnetic depending on
the Cu-O-Cu bond angle \cite{Mizuno}. The classical ground state of
this model is a helix whose pitch angle $\phi_b$ is
given by the expression

\begin{equation}
\cos \phi_b \simeq - \frac{J_1-3J_3}{4(J_2-3J_4)} \, ,
\label{pitch}
\end{equation}
which is valid as long as $|J_1|$, $|J_3|$ and $|J_4|$ are
significantly smaller than $|J_2|$. Using a finite temperature Lanczos
technique \cite{Jaklic}, we have calculated the temperature dependent
susceptibility on chains of length $N=24$ for models where condition
(\ref{pitch}) is fulfilled, using the angle $\phi_b=82^\circ$ obtained
in the analysis of the neutron scattering data. An excellent fit (red
line in Fig. 2) is obtained for the entire temperature range from just
above $T_{\rm N}$ up to 330 K, with the parameters $J_1 =-16.4$ K,
$J_2 =90$ K, $J_3 =7.2$ K, and $J_4 =6.3$ K. The $g$-factor, $g=2.14$,
is typical for Cu$^{2+}$ in square-planar geometry. The fact that the
antiferromagnetic interaction $J_2$ is by far the largest parameter in
the spin Hamiltonian explains the surprisingly good fit obtained by
the nearest-neighbor model discussed above. The longer-range couplings
$J_3$, $J_4$ are small, as expected, yet they are essential in this
context, since fits attempted with $J_3 = J_4 =0$ ($J_1/J_2=-0.52$
which yields the experimental pitch angle, green line in Fig. 2) are
in significantly poorer agreement with the experimental data.

\begin{figure}
% \centerline{\includegraphics[width=8.0cm,angle=0]{LCfig4.eps}}
\centerline{\includegraphics[width=8.0cm,angle=0]{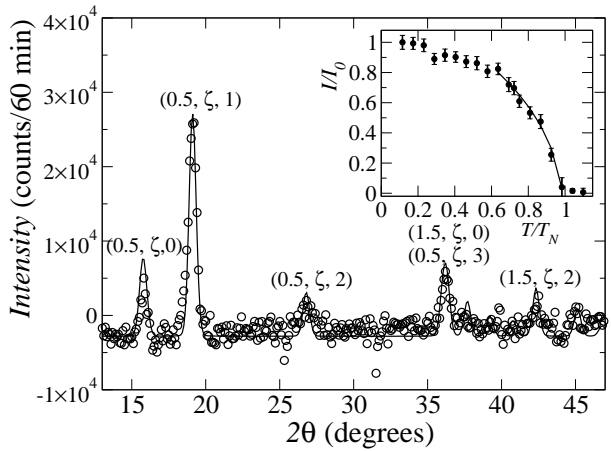}}
\caption{Difference of neutron diffraction data collected at 2K and
20K (open circles). The extra reflections at low temperature are all
of magnetic origin. The solid line is a fit to a spiral model in the
$b$$c$-plane with propagation vector $(0.5,\zeta,0)$ with
$\zeta=0.227$. The temperature dependence of the $(0.5,\zeta,1)$
reflection is plotted in the inset together with a fit of the
intensity to a power law.}
\label{LCfig4.eps}
\end{figure}

We have thus determined the magnetic ground state of NaCu$_2$O$_2$, a
spin-1/2 chain compound without significant substitutional disorder,
as a magnetic helix with a substantial ordered moment of $0.56
\mu_B$. This rules out impurity-induced magnetic long-range order, a
scenario proposed \cite{Masuda} for LiCu$_2$O$_2$, where a helix with
a similar propagation vector was observed. The spin Hamiltonian is
dominated by a large antiferromagnetic exchange parameter $J_2\sim 90$
K, which can be identified as the interaction between second-nearest
neighbors along the spin chains. The dominance of the $J_2$ coupling
may at first seem surprising, but it can be understood as a
consequence of the structural geometry of NaCu$_2$O$_2$, whose Cu-O-Cu
bond angle, $92.9^\circ$, is very close to the critical angle of
$\sim$94$^\circ$ at which the nearest-neighbor interaction is expected
to change sign \cite{Mizuno}. Hence, the nearest-neighbor coupling
$J_1$ is small. The magnitude of $J_2$ is in very good agreement with
theoretical calculations for edge-sharing copper oxide chains with
Cu-O bond parameters determined for NaCu$_2$O$_2$ \cite{Mizuno}. This
agreement is significant, because the prediction is largely
insensitive to the bond angle and hence quite robust, and because
direct measurements of this parameter in other copper oxide spin
chains yield comparable results. The inclusion of a small interchain
interaction within a chain pair ($J'$ in Fig.1b) marginally affects
the values of the intrachain exchange parameters extracted from the
susceptibility, and therefore does not change this picture qualitatively.

In view of our work on a clean, isostructural compound and the
theoretical work of Ref. \cite{Mizuno}, the exchange parameters of
LiCu$_2$O$_2$ should be reexamined. Because of the smaller Cu-O-Cu
bond angle, the magnitude of the nearest-neighbor coupling in the spin
Hamiltonian for LiCu$_2$O$_2$ is expected to be larger than that of
NaCu$_2$O$_2$, but the next-nearest-neighbor coupling should be
comparable. A scenario in which an antiferromagnetic interchain
interaction of magnitude $J^\prime\sim 70$ K significantly exceeds
all of the relevant intrachain interactions \cite{Masuda} appears very
unlikely \cite{Drechsler}.

In conclusion, the magnetic helix state and paramagnetic
susceptibility of NaCu$_2$O$_2$ are in good quantitative agreement
with the predictions of a model including longer-range exchange
interactions. An open question concerns our use of the classical
expression (\ref{pitch}) for the pitch angle. Some theoretical work
suggests that the pitch angle for quantum models deviates
substantially from this expression \cite{Bursill}. However, one has to
keep in mind that the magnitude of the ordered moment per magnetic
copper site is $ 0.56 \mu_B$, somewhat larger than the values typically
observed in 1-D systems \cite{Kojima}. Quantum zero-point
fluctuations thus appear to be significantly suppressed in the ordered
state. Large ordered moments were also observed in other
corner-sharing copper oxide spin chain compounds, and were ascribed to
large exchange anisotropies. A quantitative understanding of the
interplay between spin anisotropies and frustration in NaCu$_2$O$_2$
and other spin chain materials is an interesting subject of further
investigation.

We would like to  thank A. Ivanov, G.J. McIntyre, T. Hansen and
B. Ouladdiaf (ILL) and G. Siegle and E. Br\"{u}cher (MPI) for
experimental assistance and M. Horvatic for insightful discussions.

\bibliographystyle{prsty}

\begin{thebibliography}{99}
\bibitem{comment} Further work is required to obtain
a quantitative description of the anisotropic susceptibility of both LiCu$_2$O$_2$ \cite{Masuda} and NaCu$_2$O$_2$ (Fig. 3a) below $T_N$, which may be influenced
by a subtle spin canting that does not significantly affect the
neutron diffraction pattern.
\end{thebibliography}

\end{document}